\documentclass{emulateapj}

\newcommand{\Msun}{{\rm M_{\odot}}}

\newcommand{\kpc}{\, {\rm kpc}}

\slugcomment{}

\shorttitle{Dynamically Driven ISM Evolution in M51}
\shortauthors{Koda et al.}

\begin{document}

\title{Dynamically Driven Evolution of the Interstellar Medium in M51}

\author{Jin Koda\altaffilmark{1,2},
Nick Scoville\altaffilmark{1},
Tsuyoshi Sawada\altaffilmark{3},
Misty A. La Vigne\altaffilmark{4},
Stuart N. Vogel\altaffilmark{4},
Ashley E. Potts\altaffilmark{1},
John M. Carpenter\altaffilmark{1},
Stuartt A. Corder\altaffilmark{1},
Melvyn C. H. Wright\altaffilmark{5},
Stephen M. White\altaffilmark{4},
B. Ashley Zauderer\altaffilmark{4},
Jenny Patience\altaffilmark{1},
Anneila I. Sargent\altaffilmark{1},
Douglas C. ÐJ. Bock\altaffilmark{6},
David Hawkins\altaffilmark{7},
Mark Hodges\altaffilmark{7},
Athol Kemball\altaffilmark{8},
James W. Lamb\altaffilmark{7},
Richard L. Plambeck\altaffilmark{5},
Marc W. Pound\altaffilmark{4},
Stephen L. Scott\altaffilmark{7},
Peter Teuben\altaffilmark{4},
David P. Woody\altaffilmark{7}}

\email{jin.koda@stonybrook.edu}

\altaffiltext{1}{Department of Astronomy, California Institute of Technology, Pasadena, CA 91125}
\altaffiltext{2}{Current address: Department of Physics and Astronomy, SUNY Stony Brook, Stony Brook, NY 11794-3800}
\altaffiltext{3}{Nobeyama Radio Observatory, National Astronomical Observatory, Nobeyama, Minamimaki, Minamisaku, Nagano, 384-1305, Japan}
\altaffiltext{4}{Department of Astronomy, University of Maryland, College Park, MD 20742}
\altaffiltext{5}{Department of Astronomy and Radio Astronomy Laboratory, University of California, Berkeley, CA 97420}
\altaffiltext{6}{Combined Array for Research in Millimeter-wave Astronomy, P. O. Box 968, Big Pine, CA 93513}
\altaffiltext{7}{Owens Valley Radio Observatory, California Institute of Technology, P. O. Box 968, Big Pine, CA 93513}
\altaffiltext{8}{National Center for Supercomputing Applications, University of Illinois at Urbana-Champaign, Champaign, IL 61820}

\begin{abstract}
Massive star formation occurs in Giant Molecular Clouds (GMCs); an understanding of the evolution
of GMCs is a prerequisite to develop theories of star formation and galaxy evolution.
We report the highest-fidelity observations of the grand-design spiral galaxy M51 in carbon monoxide
(CO) emission, revealing the evolution of GMCs vis-a-vis the large-scale
galactic structure and dynamics.
The most massive GMCs (Giant Molecular Associations - GMAs) are first assembled and then broken up
as the gas flow through the spiral arms. The GMAs and their $\rm H_2$ molecules are not fully dissociated
into atomic gas as predicted in stellar feedback scenarios, but are fragmented into smaller GMCs
upon leaving the spiral arms. The remnants of GMAs are detected as the chains of GMCs that emerge from
the spiral arms into interarm regions. The kinematic shear within the spiral arms is sufficient
to unbind the GMAs against self-gravity. We conclude that the evolution of GMCs
is driven by large-scale galactic dynamics --their coagulation into GMAs is due to spiral arm streaming 
motions upon entering the arms, followed by fragmentation due to shear as they leave the arms
on the downstream side. In M51, the majority of the gas
remains molecular from arm entry through the inter-arm region and into the next spiral arm passage.
\end{abstract}


\section{Introduction}

Despite numerous studies of molecular gas in the Milky Way and galaxies \citep{sco87,bli07},
the processes affecting evolution of GMCs have remained poorly understood. In fact, uncertainty
remains as to whether GMCs are stable structures that survive over a galactic rotation period
\citep[$>10^8$ yrs; ][]{sco79, sco04} or are transient structures, destroyed immediately after 
formation by violent feedback from young stars
[a few $\times 10^7$ yrs; \citealt{bli80, elm07}, or even shorter $\sim 10^6$ yrs; \citealt{elm00, har01}].

The internal structure of GMCs is generally believed to be determined by an approximate balance of self-gravity
and internal turbulent or magneto-turbulent pressures \citep[exceeding the external ISM pressures
by two orders of magnitude; ][]{mye78}.
Therefore, GMCs are likely self-gravitationally bound, distinct, and long-lived objects \citep{sco87}.
A difficulty, however, arises in maintaining the internal turbulence over the long lifetimes,
since the turbulence should dissipate rapidly within a cloud crossing time ($\sim$a few $\times 10^{6}\,\rm yrs$).
The energy must be continuously resupplied if their lifetimes
are longer  \citep[e.g. ][]{hei01}. The energy source is still unknown although supernovae and galactic rotation
have been suggested  \citep{mac04, wad02, kod06}. Models for rapid GMC formation
and disruption have been proposed to avoid the need to re-energize the turbulence \citep{bal99}.
The absence of older $>5\,\rm Myr$ stellar populations within GMCs has been cited as evidence
of short GMC lifetimes \citep{har01}. However, more recent observations do find some older
star populations \citep{jef07, gan08, oli09}, presumably indicating longer lifetimes.
Moreover, at some point the older stars must drift free of the GMCs since they are
no longer subject to the hydrodynamic forces of the ISM gas.
The collapse timescale of star-forming cores via ambipolar diffusion is substantial,
and may support the longevity \citep{tas04}.

This controversy carries over when considering the galactic spatial  distribution of GMCs, 
since early molecular line surveys of the Milky Way linked the GMC evolution
to the Galactic disk dynamics.
\citet{coh80} argued that  GMCs were confined  largely to the spiral
arms with very few seen in the inter-arm regions, implying that GMCs must be short-lived
with a lifetime similar to the arm-crossing timescale \citep[a few $ \times 10^7$ yr;][]{dam01}.
However, \citet{san85} claimed to find many GMCs in the inter-arm regions, suggesting that the GMCs must
be quasi-permanent structures, surviving $\gtrsim10^8$ yrs, i.e. a substantial fraction of a galactic
rotation period. A recent $^{13}$CO survey suggests that GMCs in the inter-arm
regions are less-massive \citep{kod06, jac06}, accounting for the earlier discrepancies if the Cohen et al.
survey missed the less-massive inter-arm GMCs. In any case, all Galactic surveys have
the intrinsic limitation that the velocity dispersion and streaming motions of the clouds can blur out
the spiral arm/inter-arm definition when the disk is viewed edge-on.

High-resolution molecular gas
observations of external face-on spiral galaxies are therefore essential for the full census of GMC
population over galactic disks. However, prior interferometers, which are required for such
high-resolution imaging, had only a small number of telescopes, and thus were severely limited
by low image-fidelity. Indeed, the high side-lobes of bright spiral arms due to poor $uv$-coverage
have often led to false structures in the inter-arm regions \citep{ran90, aal99, hel03}.

\section{Observations and Data Reduction}

High-fidelity imaging of nearby galaxies at millimetre-wavelengths has now become feasible
with the Combined Array for Research in Millimeter Astronomy  (CARMA). CARMA is
a new interferometer, combining the six 10-meter antennas of the Owens Valley Radio Observatory (OVRO)
millimeter interferometer and the nine 6-meter antennas of the Berkeley-Illinois-Maryland Association (BIMA)
interferometer. The increase to 105 baselines (from 15 and 45 respectively) enables the highest fidelity
imaging ever achieved at millimeter wavelengths. The entire optical disk of the Whirlpool galaxy M51
($6.0' \times 8.4'$) was mosaiced in 151 pointings with Nyquist sampling of the 10m antenna beam
(FWHM of 1 arcmin for the 115GHz CO J=1-0 line). The data were reduced and calibrated using
the Multichannel Image Reconstruction, Image Analysis, and Display (MIRIAD) software package \citep{sau95}.

We also obtained total power and short spacing data with the 25-Beam Array Receiver System (BEARS)
on the Nobeyama Radio Observatory 45m telescope (NRO45, FWHM = $15\arcsec$).
Using the On-The-Fly observing mode \citep{saw08}, the data were oversampled on
a 5" lattice and then re-gridded with a spheroidal smoothing function, resulting in a final
resolution of $22\arcsec$. We used the NOSTAR data reduction package developed at
the Nobeyama observatory.
We constructed visibilities by de-convolving the NRO45 maps with the beam function
(i.e. a convolution of the 15" Gaussian and spheroidal function), and Fourier-transforming
them to the $uv$-space. We combined the CARMA and NRO45 data in Fourier space,
inverted the $uv$ data using theoretical noise and uniform weighting, and CLEANed the maps
to yield a three-dimensional image cube (Right Ascension, Declination, and LSR Doppler velocity).

The combined data have an RMS sensitivity of 40 mJy/beam in 5.1 km s$^{-1}$ wide channels,
corresponding to $1 \times 10^5 \Msun$ at the distance of 8.2 Mpc (adopting a CO-to-$\rm H_2$ conversion
factor of $X_{\rm CO}=2 \times 10^{20}\,\rm cm^{-2}$ [K km s$^{-1}$]$^{-1}$). Typical GMCs in the Milky Way
\citep[i.e. $4 \times 10^5 \Msun$ in mass and 40 pc in diameter, ][]{sco87} are therefore detected at $4\sigma$ significance.
Our angular resolution of $4\arcsec$ corresponds to 160 pc, which is high enough to isolate 
(but not resolve) the GMCs,
given that the typical separation of Galactic GMCs is a few 100 pc to kpc \citep{sco87, kod06}.
The combination of spatial resolution, sensitivity, and image-fidelity differentiates our study from
previous work \citep{vog88, gar93, nak94, aal99, hel03}, and enables a reliable
census of GMCs in M51.

Figure \ref{fig:map}a shows the CO map integrated over all velocities. The new image shows the full distribution
of molecular gas over the entire optical disk of M51 ($14.3 \times 20.0\,\rm kpc^2$),
including both the prominent spiral arms and inter-arm regions.
The bright inner arms have been previously imaged \citep{vog88,aal99}, but the new data extend
these arms over the full disk and most importantly, yield significant detection of inter-arm GMCs
for the first time. Figure \ref{fig:gmc}a shows the distribution of discrete GMCs measured using CLUMPFIND
\citep{wil94}, clearly indicating many GMCs with mass exceeding $4\times10^5 \Msun$
in the inter-arm regions. Lower mass GMCs would not be detected by the cloud finding algorithm and
only 36\% of the inter-arm CO emission is seen in the detected discrete clouds shown in Figure 2.

\section{Discussion}
Two possibilities for GMC formation and lifetime can clearly be distinguished from the observed
GMC distribution relative to the spiral arms. One possibility is that they are transient, short-lived
structures -- formed locally at convergence locations of the galactic hydrodynamic flows and destroyed quickly
by disruptive feedback from star formation within the GMCs. Alternatively, if the GMCs are long-lived, 
lasting through the inter-arm crossing period,  their presence in the  inter-arm regions is naturally explained. 
A simple calculation can be done to rule out the formation of abundant GMCs from the ambient gas with
a low average-density in the inter-arm regions. Assuming a very ideal
spherical gas accretion of the converging velocity $v$ into the volume with the radius $r$, 
the mass accumulates within the time $t$ is $M = 4 \pi r^2 v m_{\rm H} n t$, where $n$
is the number density of ambient gas and $m_{\rm H}$ is the mass of hydrogen atom.
Using a GMC radius $r \sim 20\,\rm pc$, typical velocity dispersion in galactic disk
$v \sim 10\,\rm $km s$^{-1}$,
and average gas density $n \sim 1-5 \,\rm cm^{-3}$ (inferred for M51 in the areas without GMCs),
it takes $\sim 10^8$ yrs to accumulate  $4\times10^5 \Msun$.
This mass-flow argument is valid even when the ambient gas is not exactly diffuse but consists
of smaller clouds if their distribution is roughly uniform.
Therefore, the majority of the GMCs in the inter-arm regions cannot have formed there locally
on the required short timescales (i.e. inter-arm crossing timescale $\sim 10^8$ yrs).

Figure \ref{fig:gmc}a also reveals that the molecular gas properties, specifically their masses, are significantly
dependent on galactic environment. The most massive structures with $10^{7-8}\Msun$,
referred to as GMAs, are found only in spiral arms, not in inter-arm regions.
The improved image fidelity was necessary to avoid misidentification of even such massive GMAs,
especially in the inter-arm regions.
GMAs must therefore form and disrupt while crossing the spiral arms (with a timescale typically $\sim 2$-$5\times 10^7$ yrs).
Their formation within the arms is aided by the spiral streaming which causes deflection and
convergence of the galactic flow streamlines in the arms.
Although stellar feedback (e.g. photo-dissociation by OB star ultraviolet radiation and supernova explosions)
is often invoked for GMC destruction \citep{lar87, wil97},
these mechanisms are not a likely cause of significant GMA dissipation given their masses
\citep[$10^{7-8} \Msun$, see Figure \ref{fig:gmc}a; ][]{wil97}.
More telling is the high mass fraction of $\rm H_2$ in the inter-arm regions (Figure \ref{fig:gmc}b,c).
Comparing the mean molecular gas surface density with that of the HI \citep{bra07},
we estimate that 70-80\% of the gas remains $\rm H_2$ within the major part of the disk ($\sim 12$ kpc; Figure \ref{fig:gmc}c).
Thus, GMAs are not significantly  dissociated into atomic or ionized gas.
We conclude that the GMAs must be fragmented into the less-massive GMCs
and the most massive of these are seen as discrete clouds in the inter-arm regions in Figure \ref{fig:gmc}a.
Figure \ref{fig:gmc}a includes only the GMCs down to the $4\sigma$ level;
the sum of their emission is 36\% of the total emission of the entire disk.
The remainder of molecular emission is presumably in less-massive GMCs,
not identified as discrete clouds at the current resolution and sensitivity.
Their distribution must be roughly uniform in the inter-arm regions
since they are resolved out at $4\arcsec$ resolution.

It is unlikely that a significant portion of the remaining emission arises
in diffuse molecular gas at low densities.
First, the critical density for collisional-excitation of CO(J=1-0) transition is
a few $\times 100 \rm cm^{-3}$ in optically thick clouds,
similar to the average density within GMCs in the Milky Way \citep{sco87}.
The critical density is even higher, $\sim 3000\,\rm cm^{-3}$, in optically thin regions.
Thus, CO emission should not be detected if the density is lower
than the densities of GMCs.
Secondly, CO molecules rapidly dissociate in the diffuse interstellar radiation field
if they are unshielded by a sufficient column of dust. 
It is often discussed that a visual extinction of only $A_{\rm V}\sim 1$ mag is sufficient
for shielding \citep{pri01}; however, this is true only at high densities
\citep[$10^3 \,\rm cm^{-3}$; ][]{van88}.  A higher $A_{\rm V}$ is necessary
at lower densities, since the rate of molecular formation must be rapid
to counterbalance the rapid dissociation. At the typically densities of a few
10$^2 \,\rm cm^{-3}$ expected for molecular gas based on Galactic studies, several mag 
of visual extinction are required and the CO emitting gas must reside in GMC-like
structures during passage trough the inter-arm areas.

\begin{figure}
\epsscale{1.2}
\plotone{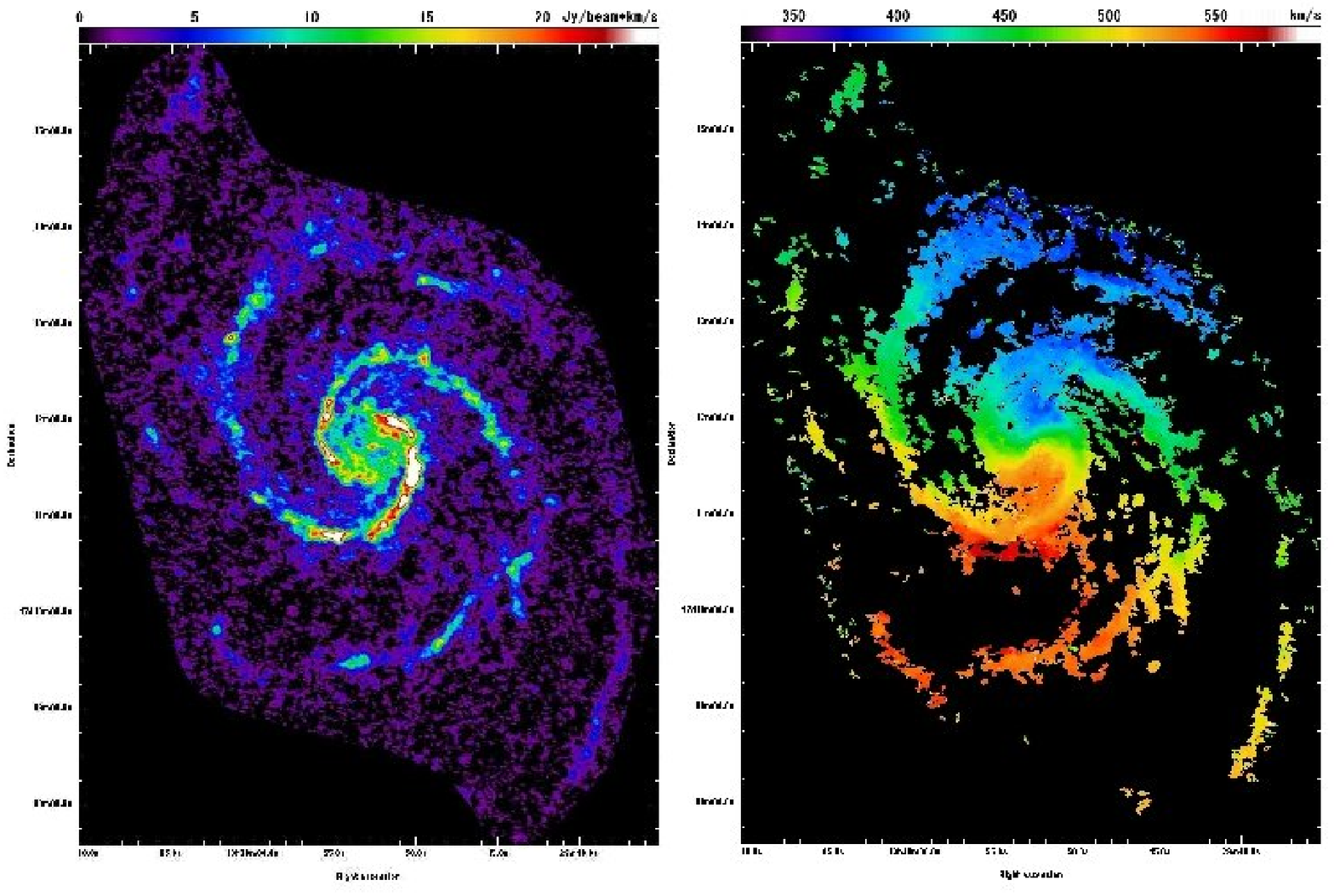}
\caption{
(a) Integrated intensity map of CO(J=1-0) emission of the entire disk of M51.
The $6.0\arcmin \times 8.4\arcmin$ region was mosaiced in 151 pointings at $4\arcsec$ resolution
with the CARMA interferometer. The total power and short-spacing data are obtained with
the On-The-Fly mapping mode of the BEARS multi-beam receiver
on the Nobeyama Radio Observatory 45m telescope (NRO45).
The CARMA and NRO45 data are combined in the Fourier space.
The map clearly detect GMCs over the entire disk for the first time,
including both the prominent spiral arms and inter-arm regions.
(b) Velocity field. Significant shear motions are seen at tangential positions
(PA of the disk kinematic major axis is -11deg)\label{fig:map}}
\end{figure}

\begin{figure}
\epsscale{1.2}
\plotone{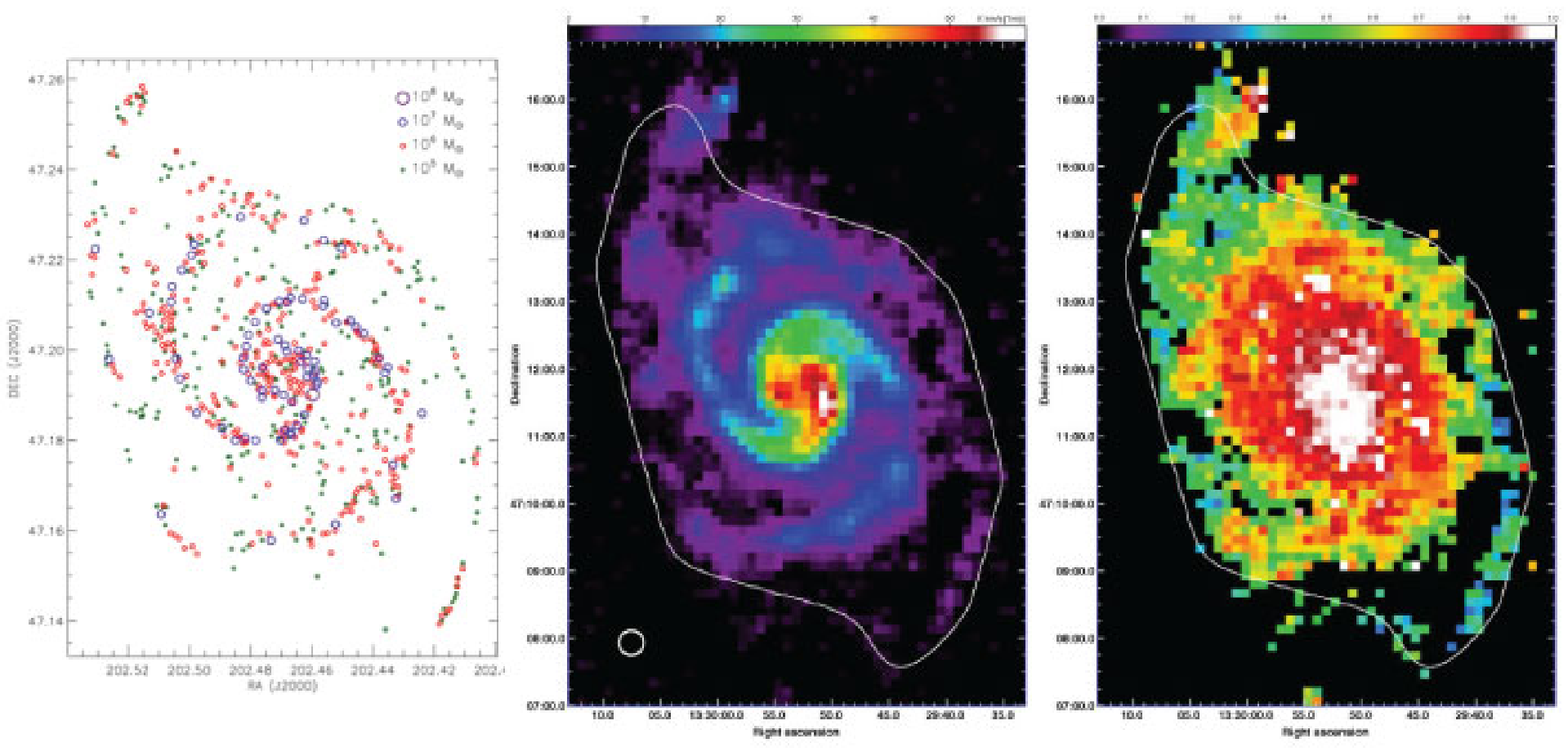}
\caption{
(a) Distribution of GMCs  ($10^{5-6} \Msun$) and giant molecular associations
(GMAs;  $>10^7 \Msun$) in M51. The GMCs/GMAs are identified with the CLUMPFIND
algorithm \citep{wil94}, down to $4\sigma$ significance, corresponding to
the typical GMC mass in the Milky Way \citep[$4 \times 10^5 \Msun$; ][]{sco87}.
The small green circles includes only the GMCs with mass above $4 \times 10^5 \Msun$.
The GMAs are seen only in the spiral arms, suggesting that they are assembled
and broken up as the gas flows through the spiral arms. Numerous GMCs are
still seen in the inter-arm regions, indicating that they survive while crossing
the inter-arm regions.
(b) Nobeyama 45 m telescope CO(J=1-0) map.
The circle at the lower-left corner is the 22$\arcsec$ beam.
The white contour around the emission indicates roughly the coverage of CARMA
observations.
(c) Molecular gas fraction defined as $2n_{\it H_2}/(n_{\rm H} + 2 n_{\rm H_2})$,
calculated with H$_{\rm I}$ data from \citet{bra07}.
The fraction is high within the major part of the disk and does not change azimuthally,
indicating that the gas stays molecular over a revolution.
\label{fig:gmc}
}
\end{figure}

Lastly, we address the nature of the GMAs: are they distinct clouds,
and simply an extension of GMC mass spectrum, or  the confusion of many GMCs
concentrated by orbit crowding in the spiral potential but unresolved
due to the limited spatial resolution.
The gas surface densities within the GMAs is $200$-$1000 \Msun \,\rm  pc^{-2}$,
generally higher than that in Galactic GMCs \citep[$\sim 170 \Msun \,\rm pc^{-2}$; ][]{sol87}.
The FWHM thickness of the molecular gas disk in M51 is unknown but in the Galaxy
it is $\sim 120$ pc \citep{sco87}, so the average gas densities within the GMAs
($40$-$200 \,\rm cm^{-3}$) are similar to typical GMC densities \citep{sol87}.
Thus, the molecular gas would continuously fill the entire volume within GMAs.
We conclude that the GMAs are most likely not just confusion of multiple GMCs,
but instead must be discrete structures. The GMCs coagulate and form GMAs in spiral arms.
Such mass concentrations would be susceptible to kinematic fragmentation
in high shear gradients across the spiral arms. In fact, the shear gradients are
$200$-$600\,\rm km \,s^{-1} \, kpc^{-1}$ (Figure \ref{fig:map}b),
large enough to pull apart the GMAs against their self-gravity;
the shear timescale, defined as the inverse of OortÕs A-constant, is comparable
to the free-fall timescale ($\sim 5$-$10$ Myr).
In addition the GMAs in our images are in virtually all instances elongated
along the spiral arms, instead of the round shapes expected if they are strongly self-gravitating.
Formation of massive GMAs without an aid of gravity is seen in  theoretical models due to
GMC agglomeration in the spiral density wave
\citep{dob06}, and due to hydrodynamic instabilities triggered by strong shear
upon entering the spiral arms \citep{wad04,wad08}.

\begin{figure}
\epsscale{.8}
\plotone{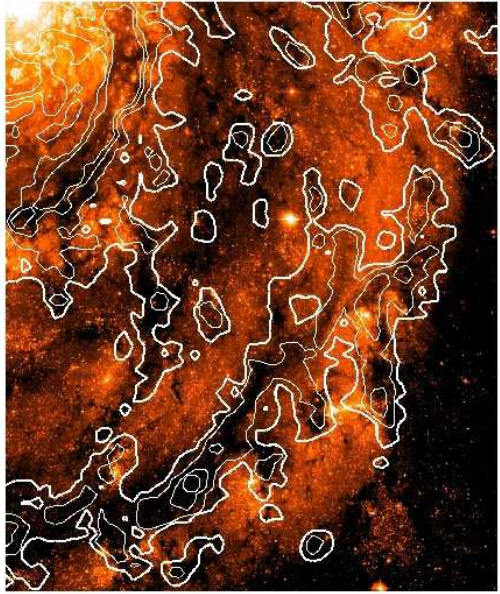}
\caption{
Spurs in inter-arm regions. Spurs that originate from the spiral arms and extend
into the inter-arm regions are seen as optical extinction in the $B$-band image from
{\it the Hubble Space Telescope (HST)} archive (color). Molecular gas (contours) traces
the spurs, possibly the fragmented remnants of GMAs.
Contours are at 3, 5, 9, 13$\sigma$ in each velocity channel,
and the lowest contour is presented with a thick line.
The spurs have very narrow line widths and are seen clearly in channel maps,
but not as much in an integrated intensity map.
We therefore overlayed the contours of all channels with emission exceeding 3 $\sigma$.
\label{fig:spurs}
}
\end{figure}

The remnants of fragmented GMAs are seen in the inter-arm regions.
Optical and infrared images show spur structures emerging from the spiral arms
into the inter-arm regions \citep{lav06} as dark filamentary lanes originating
on the outside (downstream) of spiral arms (Figure \ref{fig:spurs}). A few spurs have
also been detected in CO line \citep{cor08} and in Figure \ref{fig:spurs} here.
They are formed by fragmentation of GMAs leaving the arms into the inter-arm regions.
With the observed shear motions, GMAs would naturally shear into such  filamentary spur structures.
The total masses of spurs are a few $\times 10^{6-7} \Msun$,
approaching the masses of typical GMAs ($\sim 10^7 \Msun$).

These new observations suggest that the evolution of the dense ISM in M51
is dynamically driven. The GMCs coalesce into GMAs as they flow into
the spiral arms and the cloud orbits converge in the spiral arm.
The very massive GMA seen in the spiral arms must have lifetimes
comparable to the time needed to cross the arms ($\sim 2-5 \times 10^7$ yrs).
On leaving the spiral arms, these GMAs are fragmented by the strong shear
motions and then ejected into the downstream inter-arm regions as
lower mass GMCs. Our new observations reveal over a hundred of
the most massive GMCs ($>4 \times 10^5 \Msun$) in the interarm regions.
These GMCs are detected throughout the interm areas and therefore
must have lifetimes comparable with the interarm crossing time
of $10^8$ yrs. The majority of the interarm molecular gas is not resolved
at our current detection limit $\sim 4 \times 10^5 \Msun$ -- thus the fragmentation
of GMAs must proceed to even lower mass GMCs.
It is clear that most of the molecular gas is converted not simply to HI but to GMCs,
since the molecules dominate the overall gas abundance in the major part
of the disk (central $\sim 12 \kpc$).
The lifetimes of the lower mass GMCs require more sensitive observations.


\acknowledgments
We thank an anonymous referee for thoughtful comments.
Support for CARMA construction was derived from the Gordon and Betty 
Moore Foundation, the Eileen and Kenneth Norris Foundation, the Caltech 
Associates, the states of California, Illinois, and Maryland, and the 
National Science Foundation. Ongoing CARMA development and operations 
are supported by the National Science Foundation under a cooperative 
agreement, and by the CARMA partner universities.
This research is partially supported by HST-AR-11261.01.




\end{document}